\documentclass[final]{svjour2}
\usepackage{graphicx}
\usepackage{rotating}
\usepackage{amssymb}
\usepackage{mathptmx}
\usepackage{color}
\usepackage{amsmath} 
\makeatletter
\journalname{Journal of Low Temperature Physics}

\begin{document}

\newcommand{\hdblarrow}{H\makebox[0.9ex][l]{$\downdownarrows$}-}
\title{Fermions in Two Dimensions: Scattering and Many-Body Properties}

\author{Alexander Galea$^1$ \and Tash Zielinski$^1$ \and \\ Stefano Gandolfi$^2$ \and Alexandros Gezerlis$^1$}

\institute{$^1$ Department of Physics, University of Guelph,\\ \hspace{1cm}Guelph, ON N1G 2W1, Canada\\
\\
$^2$ Theoretical Division, Los Alamos National Laboratory,\\ Los Alamos, NM 87545, USA\\}

\maketitle

\begin{abstract}
Ultracold atomic Fermi gases in two-dimensions (2D) are an increasingly popular topic of research. The interaction strength between spin-up and spin-down particles in two-component Fermi gases can be tuned in experiments, allowing for a strongly interacting regime where the gas properties are yet to be fully understood. We have probed this regime for 2D Fermi gases by performing T = 0 {\it ab initio} diffusion Monte Carlo (DMC) calculations. The many-body dynamics are largely dependent on the two-body interactions, therefore we start with an in-depth look at scattering theory in 2D. We show the partial-wave expansion and its relation to the scattering length and effective range. Then we discuss our numerical methods for determining these scattering parameters. We close out this discussion by illustrating the details of bound states in 2D. Transitioning to the many-body system, we use variationally optimized wave functions to calculate ground-state properties of the gas over a range of interaction strengths. We show results for the energy per particle and parametrize an equation of state. We then proceed to determine the chemical potential for the strongly interacting gas.

\keywords{Cold Atoms, Fermions, Two-Dimensional Systems, Scattering, Quantum Monte Carlo}

\end{abstract}

\section{Introduction}

In recent years cold atomic gas experiments have seen novel developments\cite{Bloch:2008,Giorgini:2008,Levinsen:2015}. With the advent of Feshbach resonances it is now possible to probe the interactions of these systems in regimes ranging from tightly bound Bose-Einstein Condensate (BEC) dimers to weakly interacting Bardeen-Cooper-Schrieffer (BCS) pairs. This has allowed for the possibility of experimental verification of ground state properties, providing a strong motivation for further investigation into these cold dilute gas systems. The BEC-BCS crossover of cold atomic gases is of particular interest due to the existence of a scale independent unitary regime. In application to the crossover, mean-field theory provides a quantitatively inadequate description, which is, however, useful as a guidepost. As a result, the unitary regime has been the target of several 
first-principles Quantum Monte Carlo attempts to determine the ground state properties of these dilute cold atomic gas systems
\cite{Carlson:2003,Chang:2004,Astrakharchik:2004,Forbes:2011,Gandolfi:2011,Forbes:2012}. Intriguingly, BEC-BCS crossover of cold atomic gases is closely related to the physics of neutron matter in compact stars, which is found on the BCS side of the 
crossover. \cite{Gezerlis:2008,Carlson:2012,Stein:2012,Gandolfi:2015,Buraczynski:2016,Lacroix:2017}. 

A rich area of study subject to ongoing investigation is that of  low dimensionality \cite{Gunter:2005,Liu:2010,Martiyanov:2010,Valiente:2011,Frohlich:2011,Feld:2011,Orel:2011,Makhalov:2014,Bauer:2014,Mulkerin:2015,He:2015,Klawunn:2016,Anderson:2015,He:2016,Wong:2015,Murthy:2015,Ries:2015,Fenech:2016,Boettcher:2016,Rammelmuller:2015,Martiyanov:2016,Cheng:2016,Luciuk:2017}. These dilute cold atomic gases have been trapped using anisotropic potentials, resulting in a quasi-2D pancake-shape gas cloud. Very recent experiments have used box potentials to directly probe the physics of homogeneous two-dimensional Fermi gases~\cite{Hueck:2017}. These systems of reduced dimensionality display properties that are distinct from the analogous in 3D phenomena. As will be discussed in detail
in the following section, the main difference arises due to the logarithmic dependence on the coupling that appears in 2D.
Mean-field theory BCS has also been applied to 2D \cite{Miyake:1983,Randeria:1990}. As in 3D, the 2D regime is not well described by the two-dimensional BCS theory in the crossover region. As a result, 
the determination of ground-state properties of strongly interacting Fermi gases has been attempted with Quantum Monte Carlo
methods starting with a pioneering calculation using DMC \cite{Bertaina:2011}, which was later updated using Auxiliary-Field
Quantum Monte Carlo (AFQMC) \cite{Shi:2015}, as well as DMC using a more sophisticated wave function that included several variational parameters \cite{Galea:2016}. These previous works focused on the determination of the contact parameter and ground state energies throughout the crossover. 

In this paper we 
start with the two-body interaction from which our effective range and scattering length are determined. This is of particular importance to low-energy scattering phenomena, and is directly applied to the many body 2D s-wave problem. In section 2 we first provide a self-contained discussion of scattering in 2D using the partial-wave expansion. Then we define the effective range and scattering length parameters and go over their determination. Finally we discuss the formation of bound states in 2D, noting the substantial differences between 2D scattering and the well-known 3D scattering phenomena. In section 3 we discuss the strongly interacting 2D Fermi gas in the BEC-BCS crossover. First, we give a detailed discussion of the pairing function used in the many-body wave function. We explicitly determine the variational parameters used in the pairing function. A brief overview of the DMC method is given, then explicit variationally optimized DMC results for a range of interactions are shown. The equation of state is fit to these DMC results and then the corresponding chemical potential is calculated.

\section{Scattering}

The two-body problem is the starting point for our many-body study. Since three-dimensional (3D) scattering is a more familiar topic, we make sure to highlight crucial differences between this theory and the 2D one. We start with a partial-wave expansion of the time-independent Schr\"odinger equation and then define relevant scattering parameters, namely the scattering length $a_{2D}$ and effective range $r_e$. Next we show our techniques for determining $a_{2D}$ and $r_e$. Finally we describe and illustrate bound states in 2D.

\subsection{Partial Wave Expansion}
Describing the interaction of two particles in their center of mass reference frame, the two-body problem is reduced to that of one body with mass $m_r$ in the presence of a potential that represents the interaction.  This $m_r$ is the reduced mass of the two-body system, and simplifies to $m/2$ if the particles have equal mass.  The problem is further simplified by taking the interaction to be spherically symmetric. In this case the potential $V({\bf r})\rightarrow V(r)$, where $r=|{\bf r}|=\sqrt{r^2_x+r^2_y}\,$ is the separation distance between particles.  The time-dependent wave functions for such a system can be written as $\psi({\bf r},t)=\psi({\bf r})e^{-iEt/\hbar}$, where $\psi({\bf r})$ is an eigenfunction of the time-independent Schr\"odinger equation:
\begin{equation}\label{eq:2dsev0}
-\frac{\hbar^2}{2m_r}\nabla^2 \psi(r,\theta)+V(r)\psi(r,\theta)=E\psi(r,\theta)\,,
\end{equation}
and the scattering energy $E$ is the eigenvalue. The second angle (usually denoted with $\phi$ in physics) which ranges from 0 to $\pi$ in 3D has no meaning in 2D. Inserting the Laplacian for 2D polar coordinates, we can rewrite Eq.~(\ref{eq:2dsev0}) as
\begin{equation}\label{eq:schrodinger}
\frac{1}{r}\frac{\partial}{\partial r}(r\frac{\partial \psi(r,\theta)}{\partial r})+\frac{1}{r^2}\frac{\partial^2 \psi(r,\theta)}{\partial \theta^2}+ k^2 \psi(r,\theta) - \frac{2m_r}{\hbar^2}V(r)\psi(r,\theta)=0\,,
\end{equation}
where $k^2 =2m_rE/\hbar^2$.  Now we perform a partial-wave expansion, separating the wave function into a sum of the product of radial and angular terms ($l$ will turn out to be the orbital angular momentum):
\begin{equation}\label{eq:partial}
\psi(r,\theta)=\sum_{l=0}^{\infty}a_l R_l(r) T_l(\theta)\,.
\end{equation}
After this substitution and some simple rearranging we find the following equation, which must be satisfied for all $l$:
\begin{equation}\label{eq:schrodingerv2}
-\bigg[k^2 - \frac{2m_r}{\hbar^2}V(r)\bigg] = \frac{1}{r}\frac{\partial}{\partial r}\bigg(r\frac{\partial R_l(r)}{\partial r}\bigg)\frac{1}{R_l(r)} + \frac{1}{r^2}\bigg(\frac{1}{T_l(\theta)}\bigg)\frac{\partial^2 T_l(\theta)}{\partial \theta^2} \,,
\end{equation}
where we have removed the summation over partial waves. This can be rewritten as
\begin{equation}\label{eq:schrodingerv2a}
f(r)+g(\theta)=0\,,
\end{equation}
where $f$ depends only on $r$ and $g$ includes all the angular dependence.  If we imagine keeping $r$ fixed, then $g(\theta)=\big(1/T_l(\theta)\big)\partial^2 T_l(\theta)/\partial \theta^2$ must be the same for all values of $\theta$ if Eq.~(\ref{eq:schrodingerv2a}) is to be consistent.  The angular functions $T_l(\theta)$, therefore, must satisfy the wave equation:
\begin{equation}\label{eq:Tv1}
\frac{\partial^2 T_l(\theta)}{\partial \theta^2}=-c_l T_l(\theta)\,,
\end{equation}
where $c_l$ is some constant (which can be different for each value of $l$). The general solution can be written as
\begin{equation}
T_l(\theta)=a_l\sin(\sqrt{c_l}\,\theta)+b_l\cos(\sqrt{c_l}\,\theta)\,.
\end{equation}
Considering, for example, an incident beam at $\theta=0$, we assume symmetric scattering and can thereby eliminate the odd $\sin$ function dependence.  Then, with the simple periodic condition $T_l(0)=T_l(2\pi)$, we can determine that $\sqrt{c_l}$ must be an integer $l$ and Eq.~(\ref{eq:Tv1}) can be rewritten as~\cite{Adhikari:1986a}:
\begin{equation}\label{eq:Tv2}
\frac{\partial^2 T_l(\theta)}{\partial \theta^2}=-l^2 T_l(\theta)\,,
\end{equation}
with the normalized solution
\begin{equation}\label{eq:Tv3}
T_l(\theta)=\frac{1}{\sqrt{\pi}}\cos(l\theta)\,.
\end{equation}
Like the Legendre polynomials involved in the analogous 3D partial wave expansion, the $T_l(\theta)$ functions are linearly independent and therefore we were justified in removing the summation for Eq.~(\ref{eq:schrodingerv2}).  Using Eq.~(\ref{eq:Tv2}) to simplify the angular term and making the substitution $R_l(r)\rightarrow u_l(r)/\sqrt{r}$ to simplify the radial term, Eq.~(\ref{eq:schrodingerv2}) becomes
\begin{equation}\label{eq:schrodingerv3}
\begin{split}
- \,\bigg[k^2 - \frac{2m_r}{\hbar^2}V(r) - \frac{l^2}{r^2}\bigg] &=\bigg(\frac{1}{u_l(r)\sqrt{r}}\bigg)\frac{\partial}{\partial r}\bigg[r~\frac{\partial}{\partial r}\Big(\frac{u_l(r)}{\sqrt{r}}\Big)\bigg] \,\\
&= \bigg(\frac{1}{u_l(r)\sqrt{r}}\bigg)\frac{\partial}{\partial r}\bigg[\sqrt{r}~\Big(\frac{\partial u_l(r)}{\partial r}-\frac{u_l(r)}{2r}\Big)\bigg] \,\\
&= \frac{1}{u_l(r)\sqrt{r}}\bigg[\frac{\partial^2 u_l(r)}{\partial r^2}\sqrt{r}+\frac{u_l(r)}{4r^{3/2}}\bigg]  \,\\
&= \frac{1}{u_l(r)}\frac{\partial^2 u_l(r)}{\partial r^2}+\frac{1}{4r^2}\,.\\
\end{split}
\end{equation}
Rearranging, we find the Schr\"odinger equation for the 2D reduced radial wave function $u_l(r)$:
\begin{equation}\label{eq:2dsev1}
-\, \frac{\partial^2 u_l(r)}{\partial r^2} = u_l(r)\bigg[k^2 - \frac{2m_r}{\hbar^2}V(r) - \frac{l^2-1/4}{r^2}\bigg]\,,
\end{equation}
Differing from the 3D case, the solution to the 2D equation is related to the radial wave function by $u_l(r)=\sqrt{r}R_l(r)$ instead of $u_l(r)=r\,R_l(r)$.

For purely s-wave scattering we consider only the $l=0$ partial wave:
\begin{equation}\label{eq:2dse}
-\frac{\partial^2 u_0(r)}{\partial r^2} = u_0(r) \bigg[k^2 - \frac{2m_r}{\hbar^2} V(r) + \frac{1}{4r^2}\bigg]\,.
\end{equation}
The wave function $u_0(r)$ can be solved for numerically, although analytic solutions exist for simple potentials such as the square well.  The singularity at $r=0$ in Eq.~(\ref{eq:2dse}) means that boundary conditions must be carefully selected. In the asymptotic region, when the radial separation is larger than the range of the potential, Eq.~(\ref{eq:2dse}) simplifies to
\begin{equation}\label{eq:asymptotic}
\frac{\partial^2 u_0(r)}{\partial r^2}+u_0(r)\bigg[k^2+\frac{1}{4r^2}\bigg]=0\,.
\end{equation}
We stress at this point the significance of the $1/r^2$ term in this equation: note that it is present even for purely
$s$-wave scattering. This is to be compared with the 3D case, where a centifugal barrier is only present for beyond-$s$-wave
partial waves. As a result of this $1/r^2$, together with the different relationships between $u_0(r)$ and $R_0(r)$, the 2D problem always supports a bound state for pairwise attractive interactions 
(regardless of the strength of the interaction), in direct contradistinction to what goes on in the 3D problem.

A solution for $u_0(r)$ can be written in terms of Bessel functions of the first ($J_n$) and second ($N_n$) kind~\cite{Khuri:2009}:
\begin{equation}
u_0(r)\propto\sqrt{r}\big[J_0(kr)\cos\delta_0-N_0(kr)\sin\delta_0\big]\,.
\end{equation}
By calculating $u_0(r)$ at two radial separations $r_1$ and $r_2$ beyond the range of the potential (always for a small $k$), the s-wave phase shifts $\delta_0$ can be calculated with the relation
\begin{equation}
\cot\delta_0=\frac{\tilde{K}N_0(kr_1)-N_0(kr_2)}{\tilde{K}J_0(kr_1)-J_0(kr_2)}\,,
\end{equation}
where $\tilde{K}=u_0(r_2) \sqrt{r_1}/u_0(r_1) \sqrt{r_2}$.

\subsection{Scattering length and effective range}\label{sec:scatteringparams}
To define the $s$-wave scattering length $a_{2D}$ and the effective range $r_e$ we first look at the zero-energy Schr\"odinger equation in the asymptotic region. We take Eq.~(\ref{eq:2dse}) outside the range of the potential and set $k=0$ to find:
\begin{equation}\label{eq:e->0}
\frac{\partial^2 y_0(r)}{\partial r^2}=-\frac{1}{4r^2}y_0(r)\,,
\end{equation}
where $y_0(r)$ represents the asymptotic form of $u_0(r)$. In general, we can write the solution $y_0(r)=\sqrt{r}(\alpha+\beta\log(r))$~\cite{Khuri:2009}.  In this and all other instances we take $\log$ to represent the natural logarithm.  The 2D scattering length $a_{\rm 2D}$, in analogy to the 3D interpretation, is defined as the $r$-intercept of $y_0(r)$.  For sufficiently short range potentials, such as those used in Section~\ref{sec:2dfermigas}, $a_{\rm 2D}$ can also be given by the $r$-intercept of $u_0(r)$ for even strongly bound states.  Using the condition $y_0(a_{\rm 2D})=0$, where $a_{\rm 2D}$ is the 2D scattering length, we set $\alpha=-\beta\log(a_{\rm 2D})$ and the solution becomes $y_0(r)=\beta\sqrt{r}\log(r/a_{\rm 2D})$.  The choice for $\beta$ will influence the effective range; we set $\beta=-1$ and use the solution
\begin{equation}\label{eq:2dy0}
y_0(r)=-\sqrt{r}\log(r/a_{\rm 2D})\,.
\end{equation}
An analogous parameter to the above $\beta$ is encountered for the 3D case, where it can be determined by setting $y_0(0)=1$.  Ideally, we would use this condition for 2D scattering as well, however $y_0(r)$ in 2D does not extrapolate to $r < 0$ and always approaches 0 at the origin. This is related to the fact that $a_{\rm 2D}$, unlike the 3D scattering length, can never be negative. In other words, a 2-particle bound state exists for even arbitrarily weak attraction.  Our choice of $\beta=-1$ for the 2D case is consistent with work done by Adhikari \textit{et al.}\cite{Adhikari:1986}.

The effective range is related to the area between $u_0(r)$ and $y_0(r)$.  In 2D this is defined by the integral~\cite{Adhikari:1986}:
\begin{equation}
\label{eq:2dre}
r_e^2 = 4 \int_{0}^{\infty}\big[y_0^2(r) - u_0^2(r)\big]_{k \rightarrow 0}\,dr \,,
\end{equation}
and is the second-order term in the effective-range expansion relating low-energy phase shifts $\delta_0$ to the scattering parameters $a_{\rm 2D}$ and $r_e$.  In 2D for small values of $k$:~\cite{Khuri:2009}
\begin{equation}
\label{eq:2dshape}
\cot \delta_0\big|_{k\rightarrow 0} \, \approx \frac{2}{\pi} \bigg[\gamma + \log\Big(\frac{ka_{\rm 2D}}{2}\Big)\bigg] + \frac{k^2 r_e^2}{4} \,,
\end{equation}
where $\gamma\approx 0.577215$ is Euler's constant.  Although differing from the 3D expansion, this relationship has the same implication; low-energy scattering is independent of the details of the potential.  The logarithmic scattering-length dependence in Eq.~(\ref{eq:2dshape}) is characteristic of 2D interactions and also appears in the asymptotic solution $y_0(r)$, Eq.~(\ref{eq:2dy0}), which describes the kinetic energy alone.

To determine $a_{2D}$ and $r_e$ for an arbitrary potential we solve for $u_0(r)$ numerically and determine the asymptotic form $y_0(r)$ by extrapolating from two points, $r_1$ and $r_2$, outside the range of the potential. We consider two generalizations of Eq.~(\ref{eq:2dy0}), and use the fact that $u_0(r)=y_0(r)$ for both $r_1$ and $r_2$ to write
\begin{equation}\label{eq:outside}
\begin{split}
u_0(r_1)=\xi \sqrt{r_1} \log(r_1/a_{\rm 2D})~~~{\rm and}~~~u_0(r_2)=\xi \sqrt{r_2} \log(r_2/a_{\rm 2D})\,,
\end{split}
\end{equation}
where $\xi$ has been introduced (in place of $\beta$) for the purposes of determining $a_{\rm 2D}$ by extrapolation.  This is done using the following equations, which we find by working from Eq.~(\ref{eq:outside}):
\begin{align}
\begin{split}
a_{\rm 2D}=r_1 \exp\Bigg[\frac{-u_0(r_1)}{\xi\sqrt{r_1}}\Bigg]\,,~~~{\rm where} ~~\xi=\frac{\sqrt{r_1}\,u_0(r_2)-\sqrt{r_2}\,u_0(r_1)}{\sqrt{r_1 r_2}\,\log(r_2/r_1)}\,.
\end{split}
\end{align}
The scattering length is then used to determine $y_0(r)$ as given in Eq.~(\ref{eq:2dy0}), which is then in turn used to scale $u_0(r)$ such that $u_0(r)=y_0(r)$ at $r_1$ and $r_2$. The effective range $r_e$ can then be determined by solving Eq.~(\ref{eq:2dre}), which depends on the difference between $y^2_0(r)$ and $u^2_0(r)$.

In order to find the correct asymptotic form $y_0(r)$, it's important to solve $u_0(r)$ up to sufficiently large $r$, outside the range of the potential.  Another technical detail has to do with the scattering energy $E=\hbar^2 k^2/2m_r$.  The parameters $a_{\rm 2D}$ and $r_e$ are defined in the limit of $E \rightarrow 0$.  To make sure that finite-energy scattering effects are not influencing our determination of the parameters, we reduce $E$ until the results for $a_{\rm 2D}$ and $r_e$ have converged.

\subsection{Bound States in 2D}\label{sec:bound}
By plotting the scattering length as a function of the potential depth, we can visualize the formation of bound states.  In this respect, a difference exists between the 2D and 3D scattering theories.  As the depth of $V(r)$ is increased, the scattering length $a_{\rm 2D}$ approaches $0$ then diverges to $+\infty$ when a new bound state is created; whereas $a_{\rm 3D}$ changes from $-\infty$ to $+\infty$ when a new bound state is formed.  In each case a scattering length of $+\infty$ corresponds to a weakly bound state that becomes tighter as scattering length decreases.  The binding energy of the particle pair in 2D is given by
\begin{equation}\label{eq:2dbind}
\epsilon_b = - \,\frac{4\hbar^2 }{m a^2_{\rm 2D} e^{2\gamma}}\,,
\end{equation}
where there would also be small correction terms for a finite effective range, but we use this zero-range expression
for concreteness.
This compares to the 3D case where, for equal mass particles, the binding energy of the two particle state is $\epsilon_{b,{\rm 3D}} = - \hbar^2 / (m a^2_{\rm 3D})$.  We note that an alternate definition of $a_{\rm 2D}$ is sometimes used in other work: $a'_{\rm 2D}=a_{\rm 2D}e^\gamma/2$, such that $\epsilon_b = - \hbar^2/(m a'^2_{\rm 2D})$.

\begin{figure}
	\centering
	\includegraphics[width=0.85\linewidth, keepaspectratio]{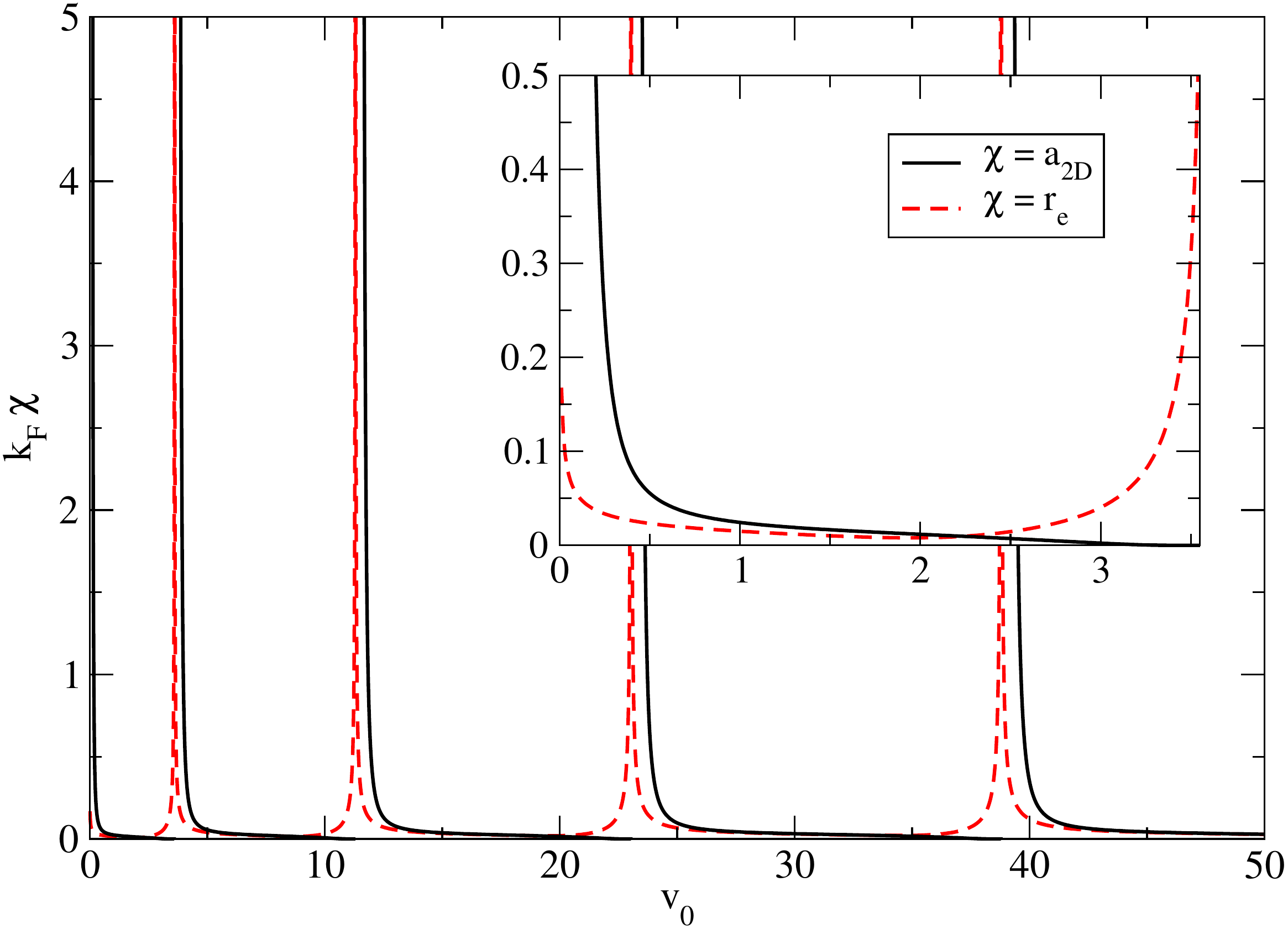}
	\caption[The scattering length (solid line, black) plotted as a function of the potential depth]{The 2D scattering length plotted as a function of the depth parameter for the potential, $v_0$.  We use the modified P\"oschl-Teller potential, Eq.~(\ref{eq:vpot}), and set $\mu/k_F=100$.  New bound states are formed when the scattering length diverges to $+\infty$ at specific values of $v_0$.  The effective range is shown with a dotted line (red), and becomes large near the locations of new bound states.  In the inset we show (in more detail) the regime where only one bound state is supported, which is $0 \leq v_0 \lesssim 3.6$ for this choice of $\mu$.}
	\label{fig:v0periodic}
\end{figure}

For the examples in this section and the DMC results presented in Section~\ref{sec:2dfermigas}, we use the modified P\"oschl-Teller potential:
\begin{equation}\label{eq:vpot}
V(r)=-v_0\frac{\hbar^2}{m_r}\frac{\mu^2}{\cosh^2(\mu r)}\,,
\end{equation}
where $r$ is the interparticle spacing. This potential is purely attractive and continuous. The parameters $v_0$ and $\mu$ roughly correspond to the depth and inverse width respectively and are tuned such that $V(r)$ reproduces the desired scattering parameters $a_{\rm 2D}$ and $r_e$.

\begin{figure}
	\centering
	\includegraphics[width=0.85\linewidth, keepaspectratio]{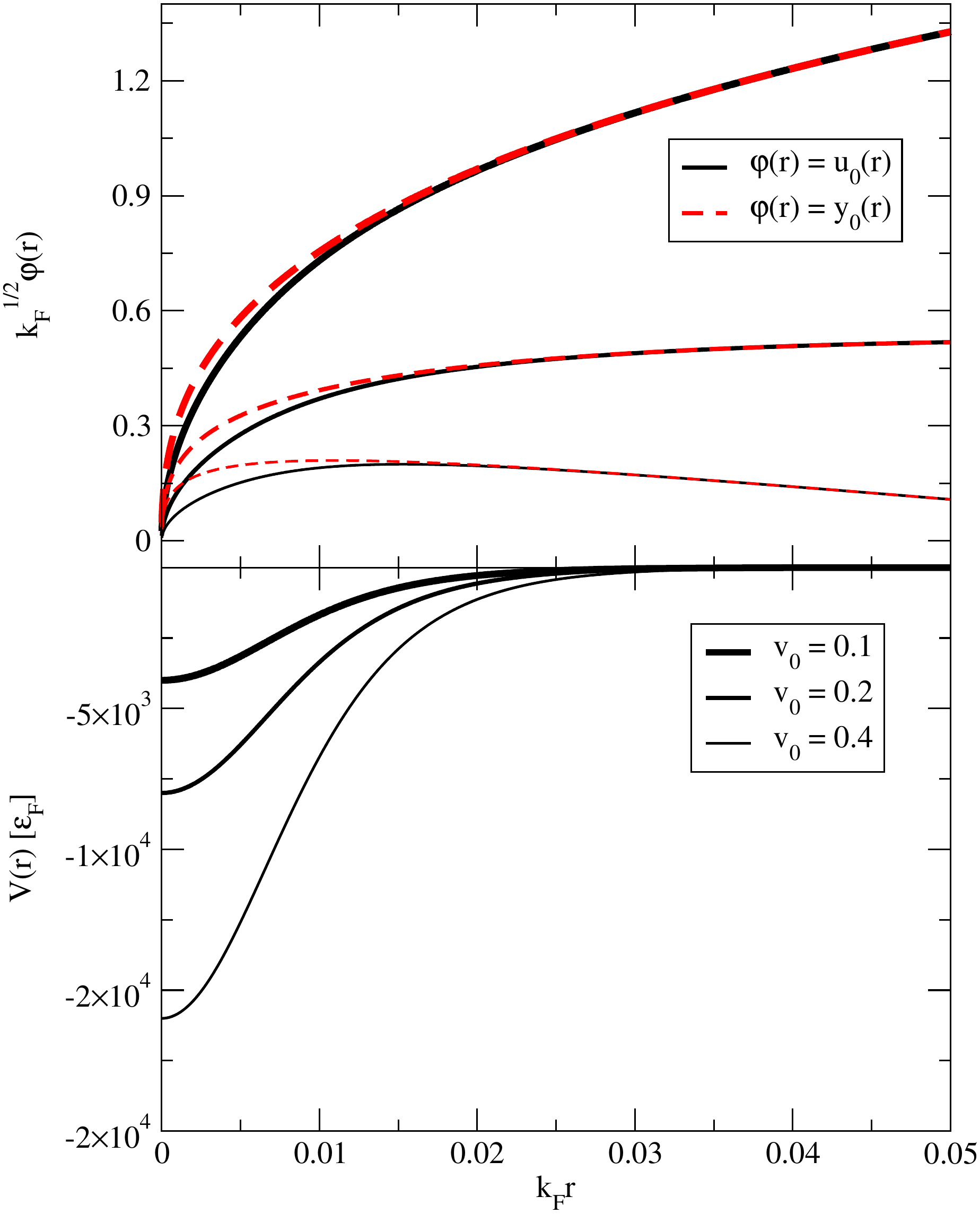}
	\caption[Reduced radial wave functions for more weakly bound states]{Reduced radial wave functions $u_0(r)$ and corresponding asymptotic solutions $y_0(r)$ are plotted with solid (black) and dashed (red) lines, respectively.  The line thickness indicates which $v_0$ value is used in the potential, Eq.~(\ref{eq:vpot}), and we set $\mu/k_F=100$.}
	\label{fig:psi1}
\end{figure}

\begin{figure}
	\centering
	\includegraphics[width=0.85\linewidth, keepaspectratio]{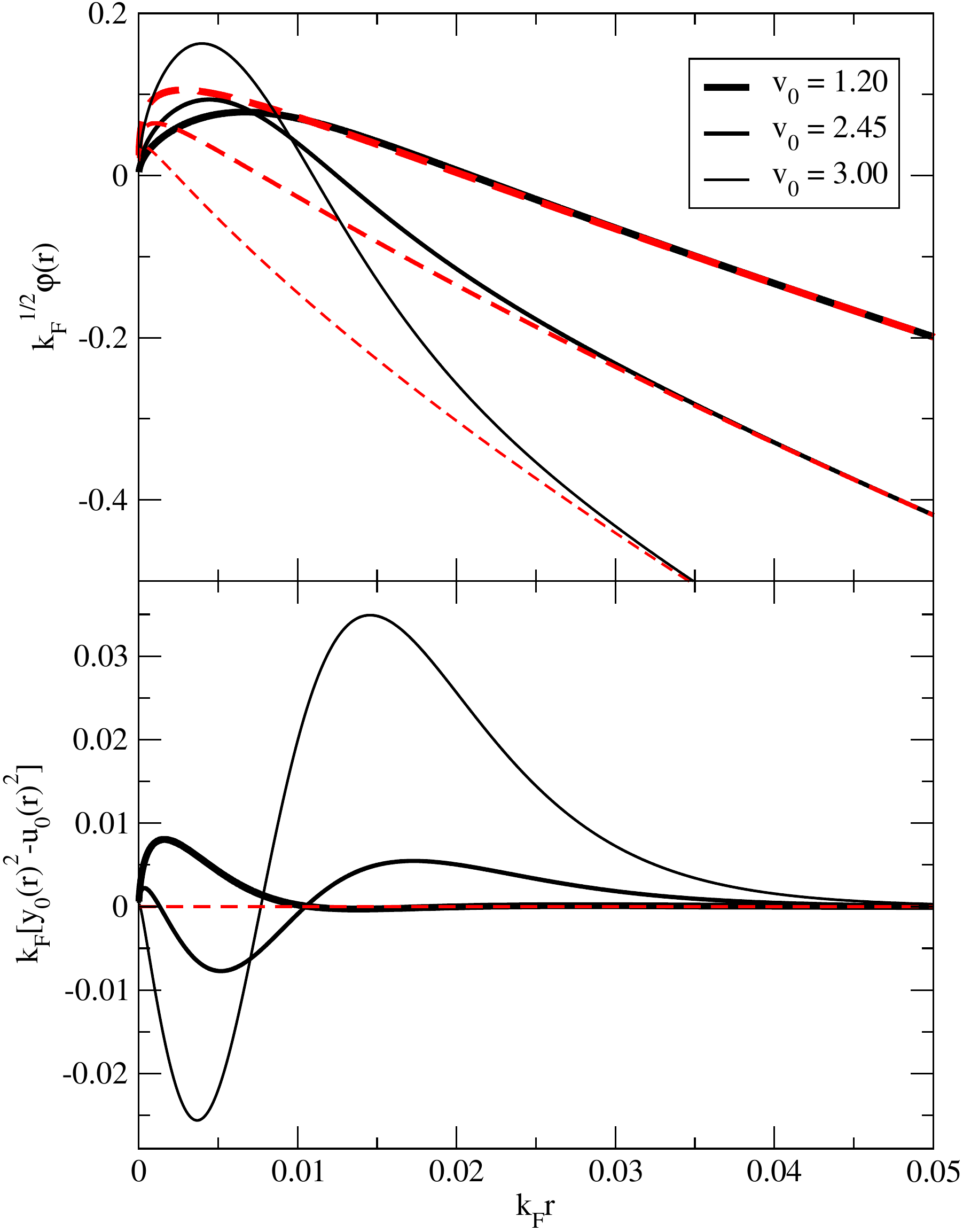}
	\caption[Reduced radial wave functions for more strongly bound states]{Reduced radial wave functions $u_0(r)$ and $v_0(r)$ are plotted in the top panel with solid (black) and dotted (red) lines.  As the depth of the potential becomes larger, approaching the regime of a second bound state, the scattering length tends towards zero and increasingly significant differences between corresponding wave functions are seen.  In the bottom panel we plot the integrand in Eq.~(\ref{eq:2dre}), where the effective range is proportional to the square root of the area under each curve.  The width parameter is fixed at $\mu/k_F=100$.}
	\label{fig:psi3}
\end{figure}

For a given $\mu$, the scattering length and effective range exhibit a repetitive pattern of spiking up and then decaying
as a function of $v_0$. This is illustrated in Fig.~\ref{fig:v0periodic}, where we have set $\mu/k_F=100$.  Here we have introduced the Fermi wave vector $k_F$ that has units of inverse length (as does $\mu$) and is related to the 2D number density $n$ by:
\begin{equation}\label{eq:kf}
k_F=\sqrt{2\pi n}\,.
\end{equation}
When $a_{\rm 2D}$ diverges to $+\infty$, a new (and initially, arbitrarily weak) bound state is created.  In other words, at these values of $v_0$, the potential becomes deep enough to support an additional bound state.  Near these locations, in this example, we see the effective range become very large.  The $v_0$ values where $a_{\rm 2D}\rightarrow\infty$ depend on the specific $\mu$ value selected.  In the case of an attractive square well potential, the effective range integral in Eq.~(\ref{eq:2dre}) will give a negative number for very strongly bound states, causing the effective range to be imaginary.  If plotting scattering parameters for the square well in the same style as Fig.~\ref{fig:v0periodic}, we would see a very similar plot.  A major distinction is that $r_e$ would continue decreasing to zero and then become imaginary as $v_0$ is increased.  Instead of gradually increasing as we approach a new bound state, $r_e^2$ becomes increasingly large and negative before diverging to a large positive value after the bound state threshold is surpassed.

\begin{figure}
	\centering
	\includegraphics[width=0.85\linewidth, keepaspectratio]{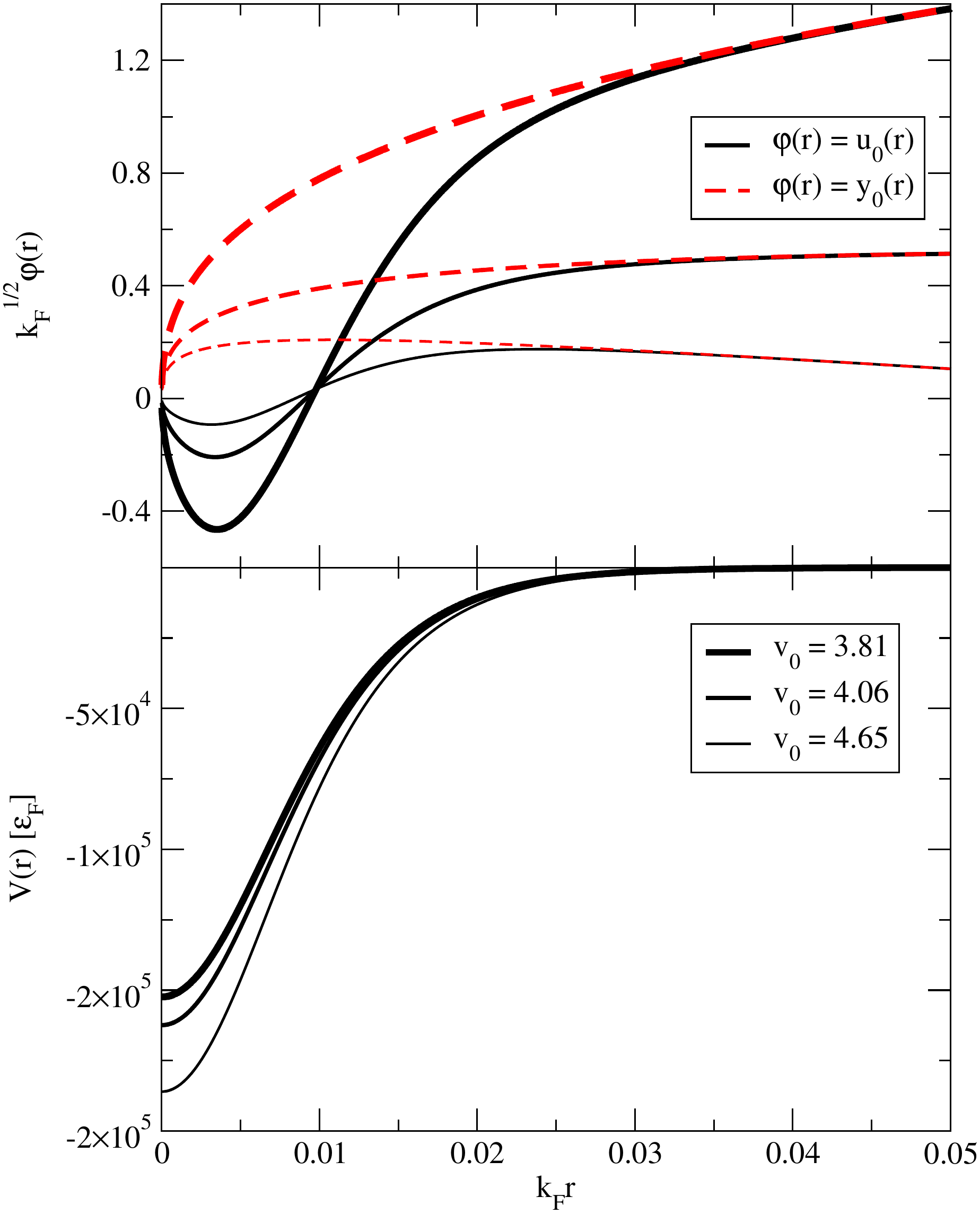}
	\caption[Reduced radial wave functions with multiple non-zero nodes]{Reduced radial wave functions are plotted in the region where two bound states can exist.  Following the style of Fig.~\ref{fig:psi1}, $u_0(r)$ is plotted with a solid line (black) and $y_0(r)$ is plotted with dashed line (red).  Line thickness represents the value of $v_0$ for the modified P\"oschl-Teller potential (plotted in the bottom panel) and the width parameter is fixed at $\mu/k_F=100$.  The extra node at $k_Fr\sim 0.01$ indicates the existence of an additional bound state and is unrelated to the scattering length.}
	\label{fig:psi2}
\end{figure}

Starting in the region where only one bound state can exist (plotted in the inset of Fig.~\ref{fig:v0periodic}), we will look at the wave function evolution as the depth of $V(r)$ is increased.  In the top panel of Fig.~\ref{fig:psi1}, we show the reduced radial wave function $u_0(r)$ (solid line) and the asymptotic form $y_0(r)$ (dashed line), as defined by Eq.~(\ref{eq:asymptotic}) in the limit as $k\rightarrow 0$ and Eq.~(\ref{eq:e->0}) respectively.  These wave functions are plotted for $v_0=0.1$, 0.2, and 0.4, with the corresponding potentials shown in the bottom panel.  We have expressed $V(r)$ in units of the Fermi energy $\epsilon_F=\hbar^2 k^2_F/2m$ and set $\mu/k_F=100$.  The 3 states in this figure are weakly bound and have large scattering lengths which are identified as the points where $y_0(r)$ would become zero.  These values are determined by extrapolating $u_0(r)$ as described in Section~\ref{sec:scatteringparams}.  In this example, the asymptotic zone is reached far before the wave functions cross the $r$-axis and therefore $u_0(a_{\rm 2D})=y_0(a_{\rm 2D})=0$.

As $v_0$ is increased further, the non-zero node of $y_0(r)$ becomes increasingly central (i.e., $a_{\rm 2D} \rightarrow 0$).  As depicted in Fig.~\ref{fig:psi3}, we find that $y_0(r)$ and $u_0(r)$ cross the $r$-axis at dramatically different locations for very strongly bound states and look far more distinct than at smaller $v_0$.  In the bottom panel, we plot the effective range integrand in Eq.~(\ref{eq:2dre}).  Here the dotted line simply marks the $r$-axis.  Curves correspond to wave functions plotted in the top panel, which can be distinguished by line thickness.  For $v_0=1.2$, the integrand is almost completely positive and peaks where the difference between $u_0(r)$ and $v_0(r)$ is maximum.  As $v_0$ is increased, the features become more pronounced and we find large negative contributions to the effective range integral.  For this example we find the positive contributions are dominant for any $v_0$.  As discussed above, this is not generally true (e.g., for the square well potential where $r_e$ becomes imaginary).

When a new bound state is formed, the scattering length diverges discontinuously to $+\infty$ and an extra node exists.  In Fig.~\ref{fig:psi2}, we show the wave function behaviour past this threshold, where the potential supports two bound states.  We plot $V(r)/\epsilon_F$ in the bottom panel as was done in Fig.~\ref{fig:psi1}.  In this figure, however, the scale has increased by an order of magnitude.  The scattering length of each state is roughly the same as $a_{\rm 2D}$ for the equivalent state in Fig.~\ref{fig:psi1}.

\section{Strongly Interacting 2D Fermi gases in the BEC-BCS Crossover}\label{sec:2dfermigas}
Now we shift to the many-body context of dilute Fermi gases with tunable interactions. We study interaction strengths where the gas is in between a BEC state of tightly bound pairs (dimers) and a weakly paired BCS superfluid.  In this regime, where the coupling of opposite-spin particles is intermediate, the gases are said to be strongly interacting and their properties are not fully understood.  As expected, we find that the mean-field BCS calculation, which gives the correct energy on each side of the crossover, is unreliable in between.

In this section we describe our many-body system including the interaction parametrizations and many-body wave function. We briefly introduce our DMC method before showing ground-state energy results for a range of interaction strengths. We first calculate the energy per particle and then parametrize an equation of state (EOS) in order to determine 
the chemical potential.

\subsection{The BEC-BCS Crossover in 2D}\label{sec:2dcrossover}

Due to the omnipresent bound state, and therefore a positive scattering length for all interaction strengths, identifying the exact region of BEC-BCS crossover point is not as obvious in 2D as in 3D. Here the crossover interaction strength is chosen to be the value at which the chemical potential switches signs; this is a reasonably intuitive choice. For $k_F a_{2D}\gg 1$ we encounter the BCS limit and for $k_F a_{2D}\ll 1$ we have the corresponding BEC limit.

Calculations are done for a range of interaction strengths, defined as
\begin{equation}\label{eq:eta}
\eta = \log(k_F a_{\rm 2D})\,,
\end{equation}
in order to determine the gas properties for a large fraction of the crossover. The number density of the many-body system is fixed such that the Fermi wave vector $k_F$, as defined in Eq.~(\ref{eq:kf}), is constant. To change the interaction strength we vary $a_{\rm 2D}$. Given that $n=N/A$ is the number density of the system (where $N$ is the number of particles and $A$ the area of the periodic box) and $r_0=1 / \sqrt{\pi n}$ is the mean interparticle spacing, the diluteness requirement is satisfied by taking $r_e \ll r_0$. We maintain a constant effective range of $k_Fr_e=0.006$ by adjusting $\mu$ as $v_0$ is varied.

\subsection{Many-Body Wave Function}\label{sec:manybody}
To describe the strongly interacting Fermi gas for any attraction strength, we use the Jastrow-BCS many-body trial wave function~\cite{Forbes:2011,Gandolfi:2011,Chang:2004,Astrakharchik:2004,Forbes:2012}:
\begin{equation}\label{eq:BCS}
\begin{split}
\Phi_{\rm BCS}({\bf R}) = {{\cal A}} [\phi({\bf r}_{11'}) \phi({\bf r}_{22'}) ... \phi({\bf r}_{N_\uparrow N'_\downarrow})]\,,\\
\Psi_T({\bf R}) = \prod_{ij'}f_{J}(r_{ij'}) \, \Phi_{\rm BCS}({\bf R}) \,,~~~~~~~~~~
\end{split}
\end{equation}
where the anti-symmetry requirement of $\Psi_T({\bf R})$ for the Fermi gas is enforced by the operator $\cal A$. Correlations between interacting particles are accounted for through the Jastrow terms $f_{J}(r_{ij'})$. The pairing functions $\phi({\bf r})$ are expressed as
\begin{equation}
\label{eq:pairfn}
\phi({\bf r}) = \sum_{n} \alpha_{n} e^{i {\bf k}_{\bf n} \cdot {\bf r}} + \tilde{\beta}(r) \,,
\end{equation}
which contains variational parameters $\alpha_{n}$ for each momentum state up to some level $n_{max}$ and the $\beta(r)$ function to account for higher-momentum contributions. This two-body function encodes details of the many-body system which vary with the interaction strength.

The spherically symmetric short-range function is given by:
\begin{gather}
\tilde{\beta}(r) = \beta(r)+\beta(L-r)-2 \beta(L/2)~~~ \mbox{for}~~ r \le L/2\,, \nonumber  \\
~~~~~~ = 0 ~~~~~~~~~~~~~~~~~~~~~~~~~~~~~~~~~~~~ \mbox{for} ~~ r > L/2\,,
\nonumber  \\
\beta (r) = [ 1 + c b  r ]\ [ 1 - e^{ - d b r }] \frac{e^{ - b r }}{d b r}~,
\label{eq:beta}
\end{gather}
which contains variational parameters $b$, $c$ and $d$.  This form of the beta function has been used for 3D calculations, and we have explicitly checked its behaviour in 2D.  Specifically, when calculating the local energy given by $\Psi^{-1}_T({\bf R}) \hat{H} \Psi_T({\bf R})$, we need to evaluate terms of the form $\partial \tilde{\beta}(r) / \partial \alpha_i$, where $\alpha_i$ is the coordinate of a specific particle (e.g., $x_2,\,y_3\ldots$) and $r=\sqrt{\Delta x^2_{ij}+\Delta y^2_{ij}}$ is the radial separation between two particles.  Defining $\Delta \alpha_{ij}$ as the projection of $r$ along a coordinate (e.g., $\Delta x_{ij}$ or $\Delta y_{ij}$) we can write:
\begin{equation}\label{eq:coordinates}
\begin{split}
\frac{\partial \tilde{\beta}(r)}{\partial \alpha_i} &= \frac{\partial \tilde{\beta}(r)}{\partial r}\,\frac{\partial r}{\partial \alpha_i}\\
&= \frac{\partial \tilde{\beta}(r)}{\partial r}\,\frac{1}{2}\frac{2 \Delta \alpha_{ij}}{\sqrt{\Delta x^2_{ij}+\Delta y^2_{ij}}}\,\frac{\partial \Delta \alpha_{ij}}{\partial \alpha_i}\\
&= \frac{\partial \tilde{\beta}(r)}{\partial r}\,\frac{\Delta \alpha_{ij}}{r}\,\frac{\partial \Delta \alpha_{ij}}{\partial \alpha_i}\\
&=\pm \frac{\partial \tilde{\beta}(r)}{\partial r}\,\frac{\Delta \alpha_{ij}}{r}\,,
\end{split}
\end{equation}
where $\partial \Delta \alpha_{ij}/\partial \alpha_i$ can be positive or negative 1.  For example, $\partial (x_2-x_5)/\partial x_2=1$ and $\partial (x_2-x_5)/\partial x_5=-1$.

The result in Eq.~(\ref{eq:coordinates}) has a singularity at $r=0$ due to the $1/r$ term.  This can cause large fluctuations in the local energy for small $r$, therefore we define the variational parameter $c$ in Eq.~(\ref{eq:beta}) such that
\begin{equation}\label{eq:findc}
\frac{\partial \tilde{\beta}(r)}{\partial r}\Big|_{r=0} = \Big[\frac{\partial\beta(r)}{\partial r}+\frac{\partial\beta(L-r)}{\partial r} - 2 \frac{\partial \beta(L/2)}{\partial r}\Big]_{r=0} = 0\,.
\end{equation}
Making use of L'H\^{o}pital's rule, it is mostly straightforward to show that
\begin{equation}\label{eq:c}
c=\frac{2+2dbL+(dbL)^2e^{bL(1+d)}+2db^2L^2e^{bL(1+d)}+2bL-2e^{dbL}bL -2e^{dbL}}{2b^2L^2(e^{dbL}-1-d+de^{bL(1+d)})}\,.
\end{equation}
In this work, we set $b=0.5k_F$ and $d=5$, as done by Gandolfi \textit{et al.}~\cite{Gandolfi:2011} for the 3D unitary Fermi gas. With these values of $b$ and $d$, we find $c \simeq 3.5$.

\subsection{DMC}\label{sec:dmc}
To determine ground-state properties of Fermi gases we use DMC to project the ground state $\Phi_0$ from the trial wave function $\Psi_T({\bf R})$. This is done by propagating in imaginary time $\tau=it$:
\begin{equation}
\begin{split}
\Phi_0=\Psi(\tau\rightarrow\infty),~~~~~~~\\
\Psi(\tau) = e^{-(\hat{H}-E_T)\tau}\Psi_T(\mathbf{R})\,,
\end{split}
\end{equation} 
where the trial energy $E_T$ is a constant offset applied to the Hamiltonian.

DMC expectation values are determined by averaging over a set of equilibrated configurations.  In this work we use the mixed estimate to calculate the energy:
\begin{equation}\label{eq:dmcener}
\begin{split}
\langle \hat{H} \rangle_M &= \frac{\langle \Psi_T | \hat{H} | \Psi(\tau) \rangle}{\langle \Psi_T | \Psi(\tau) \rangle}\\[3mm]
&= \frac{\langle \Psi_T | \hat{H}  e^{-(\hat{H}-E_T)\tau} | \Psi_T \rangle}{\langle \Psi_T | e^{-(\hat{H}-E_T)\tau} | \Psi_T \rangle}\\[3mm]
&= \frac{\langle \Psi_T | e^{-(\hat{H}-E_T)\tau/2} \hat{H} e^{-(\hat{H}-E_T)\tau/2} | \Psi_T \rangle}{\langle \Psi_T | e^{-(\hat{H}-E_T)\tau/2} e^{-(\hat{H}-E_T)\tau/2} | \Psi_T \rangle}\\[3mm]
&= \frac{\langle \Psi(\tau/2) | \hat{H} | \Psi(\tau/2) \rangle}{\langle \Psi(\tau/2) | \Psi(\tau/2) \rangle}\,.
\end{split}
\end{equation}

In the second line we wrote the explicit form of the imaginary-time evolved ket. The propagator is then used to act on the trial wave function bra straightforwardly due to the fact that it commutes with the Hamiltonian. Finally taking $\tau$ $\rightarrow$ $\infty$  we see that this is the energy of the ground state.

\subsection{Equation of State}\label{sec:eos}
A mean-field calculation~\cite{Randeria:1990} gives a ground-state energy per particle of
\begin{equation}\label{eq:meanfieldener}
E_{\rm BCS} = E_{\rm FG} + \epsilon_b/2\,,
\end{equation}
for Fermi gases in the BEC-BCS crossover, where the binding energy $\epsilon_b$ is given by Eq.~(\ref{eq:2dbind}). This is expected to be accurate for weakly paired systems in the BCS limit in which $E/N \rightarrow E_{\rm FG}$. The BEC limit of tightly bound pairs is also expected to be reasonably well described by mean field as the energy scale grows rapidly by many orders of magnitude due to large binding energies. The QMC results vary dramatically from the mean-field description in the crossover but become increasingly similar to mean-field predictions in each limit.

\begin{figure}[t]
	\centering
	\includegraphics[width=0.85\textwidth]{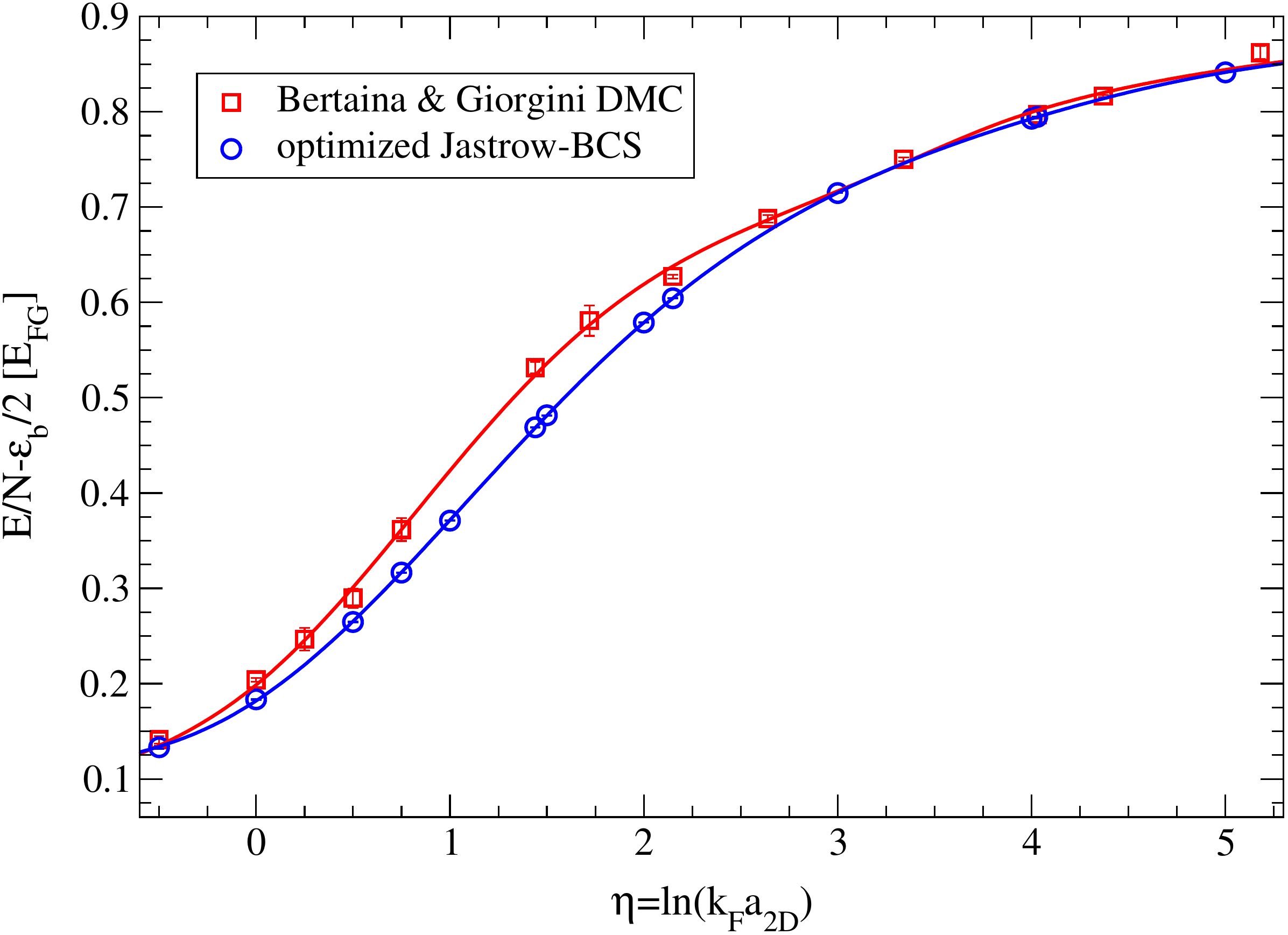}
	\caption{Our energy per particle values for 2D strongly interacting Fermi gases in the BEC-BCS crossover.  We show results of $N=26$ DMC calculations with variationally optimized Jastrow-BCS wave functions, also comparing to early DMC results.  The interaction strength is given by $\eta=\log(k_F a_{\rm 2D})$ and energies are expressed in units of the non-interacting gas $E_{\rm FG}=\epsilon_F/2$.  }
	\label{fig:ener}
\end{figure}

Our DMC calculations for a range of interactions strengths are shown in Fig..~\ref{fig:ener}. The errors represent statistical uncertainty, which becomes larger as the energy scale increases. We use the Jastrow-BCS wave function $\Phi_{\rm BCS}(\mathbf{R})$, Eq.~(\ref{eq:BCS}), which contains parameters that are optimized for each $\eta$ independently. We have compared our results to previous \textit{ab-initio} work, finding significantly lower energies than prior ground-state DMC results~\cite{Bertaina:2011} in the crossover regime and excellent agreement with AFQMC~\cite{Shi:2015}. More recently another QMC study has emerged~\cite{Madeira:2017} that finds DMC results in agreement with ours. Note that we here provide more results
in the BCS side than where available previously \cite{Galea:2016}.

In order to calculate other ground-state properties of strongly interacting Fermi gases, we calculate an EOS for our thermodynamic-limit energy results $E_{\rm TL}$. This quantity is a finite-size corrected version of the
results in Fig.~\ref{fig:ener}. The correction ranges from zero on the BEC side of the crossover to $\sim0.041~E_{\rm FG}$ in the BCS regime. Using similar methods as previous \textit{ab initio} studies~\cite{Bertaina:2011,Shi:2015,Galea:2016}, we parametrize the EOS using three functions. In the crossover regime we fit to a 7th-order polynomial:
\begin{equation}
\label{eq:7poly}
f(\eta) = \sum\limits^{7}_{i=0}c_i\eta^i\,.
\end{equation}
This is joined by a dimer form in the BEC regime~\cite{Levinsen:2015} and an expansion in $1/\eta$ in the BCS regime~\cite{Engelbrecht:1992}. The dimer form is given by:
\begin{equation}
\label{eq:BECform}
f^{\rm BEC}(\eta) = \frac{1}{2x} \bigg[1-\frac{\log(x)+d}{x}+\frac{\sum_{i=0}^2 c_i [\log(x)]^i}{x^2}\bigg]\,,
\end{equation}
where $x = \log[4\pi/(k_F a_d)^2] \approx 3.703 - 2\eta$ (for the dimer scattering length $a_d \approx 0.557a_{\rm 2D}$ \cite{Petrov:2003}) and $d = \log\pi + 2\gamma + 0.5$.  The BCS form is given by:
\begin{equation}
\label{eq:BCSform}
f^{\rm BCS} (\eta) = 1 - \frac{1}{\eta} + \sum_{i=2}^{4}\frac{c^i}{\eta^i}\,.
\end{equation}
Values of $c_i$ in Eq.~(\ref{eq:BECform}) and Eq.~(\ref{eq:BCSform}) are determined using continuity conditions for $f$, $\partial f/\partial \eta$, and $\partial^2 f/\partial \eta^2$ at the matching points. Our EOS parameters are included in Tab.~\ref{tab:eospars}. The matching points were selected as $\eta = -0.25$ and $\eta = 2.5$. We found these values result in the most optimal overall fit while including as much of the crossover polynomial function as possible.  Also, we ensure that our matching point for Eq.~(\ref{eq:BECform}) is selected on the BEC side of the crossover.

\begin{table}
	\begin{center}
		\caption[The final parameters for our EOS]{The final parameters for our EOS, where we fit to $(E_{\rm TL}/N-\epsilon_b/2)/E_{\rm FG}$.  A 7th-order polynomial $f(\eta)$ is used in the crossover regime and we fit to $f^{\rm BEC}(\eta)$ and $f^{\rm BCS}(\eta)$ on either side.  Their functional forms, Eq.~(\ref{eq:BECform}) and Eq.~(\ref{eq:BCSform}), are based on the limiting behaviour on each side of the crossover and they each contain 3 parameters that are determined by continuity restrictions.  In an effort to include as much of the intermediate polynomial as possible and minimize the overall variance of our fit, we select the matching points as $\eta = -0.25$ and $\eta = 2.5$. \label{tab:eospars}}
		\vspace{5mm}
		\begin{tabular}{c|c|c|c}
			$c_i$\, & \,$f^{\rm BEC}(\eta)$\, & \,$f(\eta)$\, & \,$f^{\rm BCS}(\eta)$ \\
			\hline
			$c_0$\, & \,28.545\, & \,0.18181\, & \, \\
			$c_1$\, & \,-42.648\, & \,0.13334\, & \\
			$c_2$\, & \,15.555\, & \,0.076788\, & \,-0.50515 \\
			$c_3$\, & \,\, & \,-0.0099012\, & \, 2.9215 \\
			$c_4$\, & \,\, & \,-0.017582\, & \, -2.8151 \\
			$c_5$\, & \,\, & \,0.0072822\, & \, \\			
			$c_6$\, & \,\, & \,-0.0010898\, & \, \\
			$c_7$\, & \,\, & \,0.000057991\, & \, \\
		\end{tabular}
	\end{center}
\end{table}

\subsection{Chemical Potential}\label{sec:chemical}
We have determined the chemical potential $\mu$ using our EOS.  In order to derive a relationship, we define
\begin{equation}
\zeta(\eta) = \frac{E_{\rm TL}(\eta)}{N}\frac{1}{E_{\rm FG}} = \frac{2E_{\rm TL}(\eta)}{N\epsilon_F}\,,
\end{equation}
where $E_{\rm TL}(\eta)$ is the total ground-state energy. This quantity is related to our parametrized EOS, where we fit to $f(\eta)=(E_{\rm TL}/N-\epsilon_b/2)/E_{\rm FG}\,$; the new quantity $\zeta(\eta)$ can easily be determined by adding $\epsilon_b/2$ in units of $E_{\rm FG}$ to $f(\eta)$. Noting that $\epsilon_F = (\pi N \hbar^2)/(mA)$ in 2D (using $k^2_F=2\pi N/A$), the chemical potential is related to $\zeta(\eta)$ as follows:
\begin{align}\nonumber
\mu = \frac{\partial E_{\rm TL}}{\partial N} &= \frac{\partial}{\partial N}\bigg[ \frac{\epsilon_F N \zeta(\eta)}{2} \bigg]\\[3mm]
\nonumber&= \frac{1}{2}\frac{\partial (\epsilon_F N)}{\partial N} \zeta(\eta) + \frac{\epsilon_F N}{2}\frac{\partial \zeta(\eta)}{\partial\eta}\frac{\partial \eta}{\partial N}\\[3mm]
\nonumber&= \frac{1}{2}\frac{\partial }{\partial N}\bigg( \frac{ \pi N^2 \hbar^2}{mA} \bigg) \zeta(\eta) + \frac{\epsilon_F N}{2}\frac{\partial \zeta(\eta)}{\partial\eta}\frac{\partial \eta}{\partial N}\\[3mm]
\nonumber&= \bigg( \frac{ \pi N \hbar^2}{mA} \bigg) \zeta(\eta) + \frac{\epsilon_F}{4}\frac{\partial \zeta(\eta)}{\partial\eta}\\[3mm]
&= \epsilon_F \zeta(\eta) + \frac{\epsilon_F}{4}\frac{\partial \zeta(\eta)}{\partial\eta}\,.
\end{align}
In the fourth line we have evaluated $\partial \eta/\partial N$ using
\begin{equation}
\frac{\partial [\log(k_F a_{\rm 2D})]}{\partial N} = \frac{\partial [\log(\sqrt{2\pi N/A}\,a_{\rm 2D})]}{\partial N} = \frac{1}{2N}\,.
\end{equation}
Expressing the chemical potential in units of the non-interacting Fermi gas, we find
\begin{equation}
\frac{\mu}{E_{\rm FG}} = 2\bigg( \zeta(\eta) + \frac{1}{4}\frac{\partial \zeta(\eta)}{\partial \eta} \bigg)\,.
\end{equation}
\begin{figure}
	\centering
	\includegraphics[width=0.85\textwidth]{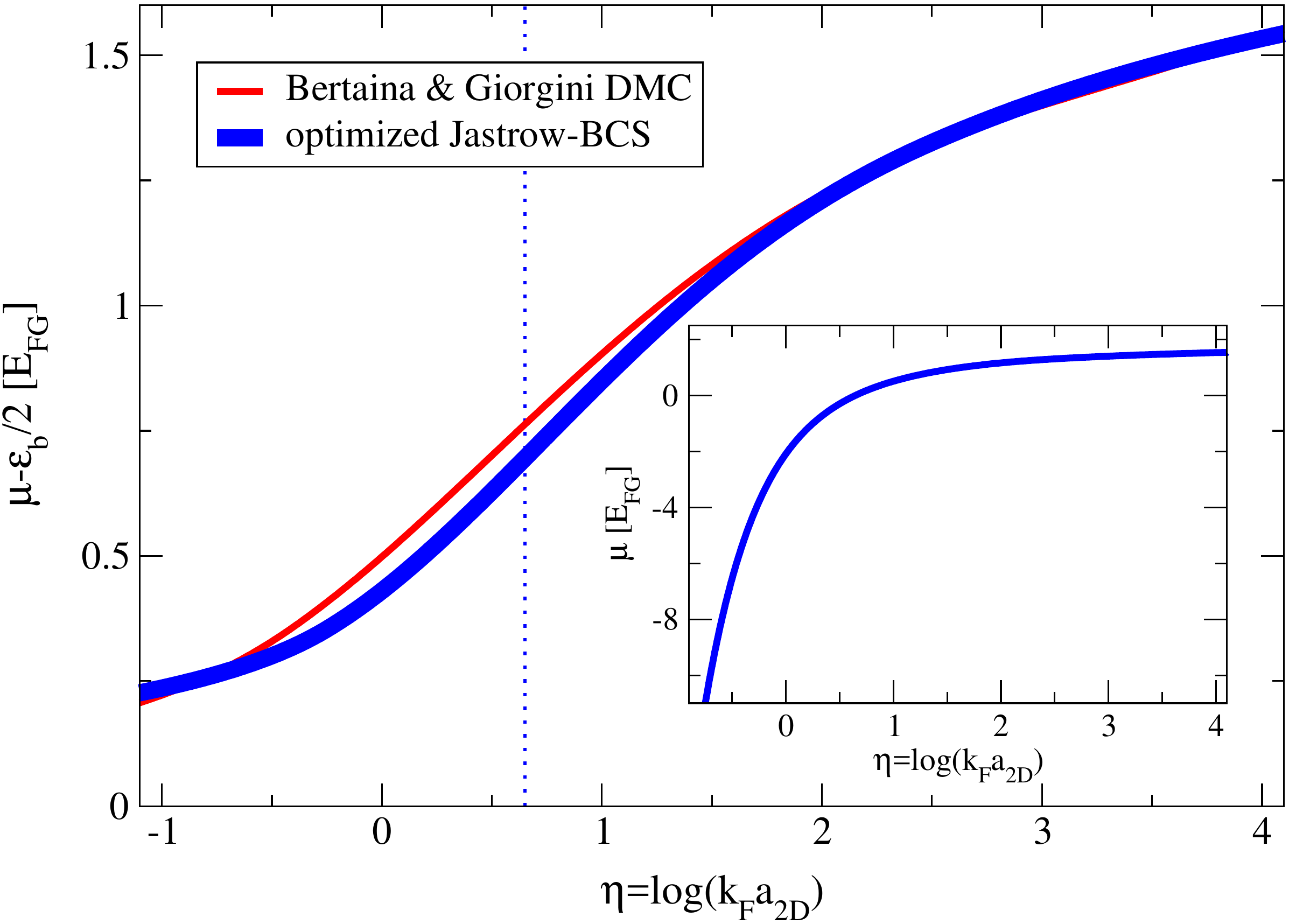}
	\caption[Chemical potential experimental comparison]{The chemical potential for strongly interacting 2D Fermi gases in the BEC-BCS crossover.  Our result (thick blue line) is shown along with a result corresponding to the Bertaina and Giorgini energies (thin red line). The inset shows the chemical potential without the binding energy offset.}
	\label{fig:chem}
\end{figure}

Our result is plotted in Fig.~\ref{fig:chem}, which shows 
the chemical potential with half of the two-body binding energy subtracted.  Our DMC determination is the 
thick blue line, which is qualitatively similar to the chemical potential one gets starting from 
the earlier DMC results \cite{Bertaina:2011} (which we generated using an analogous fitting strategy). The dotted vertical line corresponds to the interaction strength ($\eta \approx 0.65$) where the chemical potential changes sign.  
Comparing our results with experimentally extracted values of the chemical potential \cite{Boettcher:2016,Enss:2015}, 
we find a nice match in the deep BEC regime. Differences become more significant in the BCS limit, probably 
due to finite-temperature and quasi-2D effects in the experiment.

\section{Summary and Conclusion}

In this paper we presented a detailed discussion 
of scattering in the context of a two-body system confined to two dimensions. We then determined the radial wave function and defined its asymptotic form, which in a purely attractive potential always produces a bound state in contrast to the 3D case. Continuing to explore 2D scattering phenomena we illustrated the dependence of the effective range 
$r_e$ and the scattering length $a$ 
on the attractive potential's strength. Varying the potential we plotted the divergence of $r_e$ and $a$ upon the approach of new bound states. 
We also demonstrated the radial wave function and associated asymptotic form dependence on the attractive potential strength, where stronger attractive potentials produce more tightly bound states with small positive scattering lengths. 
Having quantified the 2D two body interacting system in detail, we moved to the many-body problem. Starting with the BCS determinant we looked at its composite pairing functions and specifically the variational parameters. We presented the explicit form of the $\beta$ pairing function, and briefly described the mixed-estimate method. Then we provided QMC ground-state energy results for various interaction strengths within the BEC-BCS crossover range. Fitting to these energies with a 7th-order polynomial an equation of state was determined. The expression for the chemical potential was derived according to the EOS, for which fitting parameters were explicitly provided. To conclude, the careful study of two-dimensional scattering properties in conjunction with a non-perturbative many-body method containing several variational parameters (like DMC), has led to dependable
predictions for the properties of strongly correlated physical systems.
Overall, we observe that two-dimensional strongly interacting
cold Fermi gases constitute an exciting new development, where theory can be confronted by impressive experimental work.

\begin{acknowledgements}
The authors would like to thank G. E. Astrakharchik, T. Enss, J. Thywissen, and E. Vitali 
for helpful discussions. This work was supported in part 
by the Natural Sciences and Engineering Research Council (NSERC) of Canada, the 
Canada Foundation for Innovation (CFI), the Early Researcher Award (ERA) program of the Ontario 
Ministry of Research, Innovation and Science,
the US Department of Energy,
Office of Nuclear Physics, under Contract DE-AC52-06NA25396, and the LANL LDRD program. 
Computational resources were provided by SHARCNET, NERSC, and Los Alamos Open Supercomputing.
The authors would like to acknowledge the ECT* for its warm hospitality during the ``Superfluidity and Pairing Phenomena''  workshop in March 2017, where part of this work was carried out.
\end{acknowledgements}

\pagebreak

\end{document}